\documentclass[
    ,final            
  ]
  {aipproc}

\newcommand{\apj}{{ ApJ~\/}}
\newcommand{\apjl}{{ ApJL~\/}} 
\newcommand{\apjs}{{ ApJS~\/}}

\newcommand{\mnras}{{ MNRAS~\/}}

\layoutstyle{6x9}

\begin{document}

\title{On the Origin of Metals in Some Hot White Dwarf Photospheres}


\author{Matt R.\ Burleigh}{
  address={Dept Physics and Astronomy, University of Leicester, University Road, Leicester, LE1 7RH, UK}
}

\author{Martin A.\ Barstow}{
  address={Dept Physics and Astronomy, University of Leicester, University Road, Leicester, LE1 7RH, UK}
}

\author{Jay Farihi}{
  address={Dept Physics and Astronomy, University of Leicester, University Road, Leicester, LE1 7RH, UK}
}

\author{Nigel P.\ Bannister}{
  address={Dept Physics and Astronomy, University of Leicester, University Road, Leicester, LE1 7RH, UK}
}

\author{Nathan Dickinson}{
  address={Dept Physics and Astronomy, University of Leicester, University Road, Leicester, LE1 7RH, UK}
}

\author{Paul R.\ Steele}{
  address={Dept Physics and Astronomy, University of Leicester, University Road, Leicester, LE1 7RH, UK}
}

\author{Paul D.\ Dobbie}{
  address={Australian Astronomical Observatory, PO Box 296, Epping, NSW 1710, Australia}
  }

\author{Francesca Faedi}{
  address={Astrophysics Research Centre, School of Mathematics and
  Physics, Queen's University, University Road, Belfast, BT7 1NN, UK}
}  

\author{Boris T.\ Gaensicke}{
  address={Department of Physics, University of Warwick, Coventry, CV4
  7AL, UK}
}

\begin{abstract}
We have searched for evidence for dust and gas disks at a sample of
hot DA white dwarfs $20\,000K<T_{\rm eff}<50\,000$K, without success. 
Although their atmospheres are polluted with heavy elements, 
we cannot yet convincingly and conclusively 
show that any of these objects is accreting metals from
surrounding material derived from disrupted minor planets in an old
solar system. 
\end{abstract}

\maketitle


\section{Introduction}

The discovery of a growing number of dust disks around cool ($<22\,000$K)
DAZ white dwarfs (WD) has been linked to the existence of old planetary
systems containing terrestrial bodies. The dust disks are thought to
originate from the tidal disruption of a minor planet that has
wandered too close to the WD (\citealp{jura03}). 
At temperatures $>20-25\,000$K, any dust
close to the WD will be sublimated and produce no infrared emission,
although in a few WDs $\sim20\,000$K a gas disk exists in addition to warm
dust. The gas disks are revealed by Ca\,{\sc ii} emission at $\approx8600${\AA} and
sometimes Fe\,{\sc ii} emission at $\approx5200${\AA} (\citealp{gaensicke06}).

Heavy elements are ubiquitous in the atmospheres of DA white dwarfs at
$T_{\rm eff}>50\,000$K (\citealp{barstow93,marsh97}), and are present
in some, but not all, objects at temperatures cooler than this.  Above
$50\,000$K, metals are levitated in the stellar photospheres against the
strong gravitational field by radiation pressure. For single stars, it
is impossible to distinguish whether their origin is primordial or due
to accretion. But below $\approx50\,000$K, the metal abundances sometimes
cannot be explained by diffusion theory alone. Indeed, exactly which
elements may be present, and in what abundances, can vary between
stars of seemingly identical temperatures and masses
(\citealp{barstow03}). As with
the cooler DAZ white dwarfs, the gravitational settling times for
these heavy  elements are so fast that some hot white dwarfs $<50\,000$K
may be undergoing low level accretion from an unknown external
source. The discovery of dust and gas disks around some DAZs suggests
these may also be the source of the material seen in the photospheres
of some hotter white dwarfs between $20\,000$K and $50\,000$K.

\section{Spitzer mid-infrared observations of hot white dwarfs}

We used Spitzer/IRAC to search for excess mid-IR emission,
potentially due to the presence of warm dust, in a sample of hot DA
WDs $20\,000K<T_{\rm eff}<50\,000$K, 
whose photospheric heavy element composition we
had previously studied in detail in the far-UV and EUV 
(\citealp{barstow03,bannister03}). Data were obtained in our own
programmes, and from the Spitzer public archive. Only one star,
PG$1234+482$, shows any excess emission, due to the presence of an L0
dwarf companion (\citealp{mullally07,steele07}). The
non-detection of warm dust at these hot degenerates is not entirely
unexpected, given our current expectation that the known disks lie
within a few solar radii of their WD hosts, and that at higher WD
temperatures this material will likely be sublimated.

\section{An optical search for Ca and Fe emission in hot white dwarf
  spectra}

\begin{figure}
  \includegraphics[height=.59\textheight, angle=270]{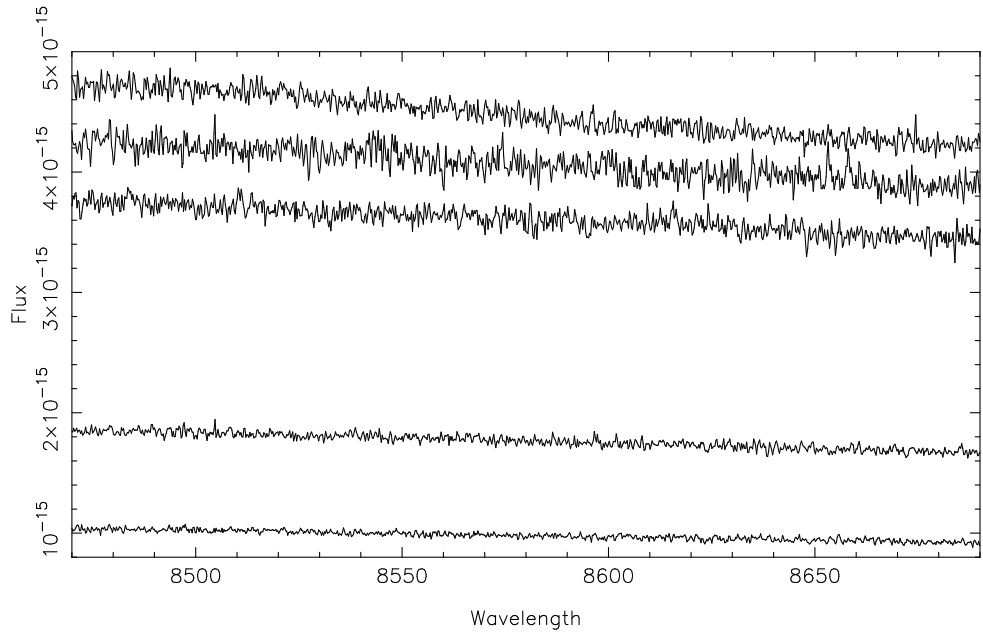}
\label{mrburleigh01}
  \caption{Red optical spectra of hot DA white dwarfs centred on the
    Ca\,{\sc ii} $8498${\AA}, $8542${\AA}, $8662${\AA} triplet. 
Top to bottom: HS$0209+083$, GD394, REJ$1943+50$, GD246, REJ$1614-085$. }
\end{figure}

We obtained optical spectra of ten hot DA white dwarfs studied by
\citet{barstow03} to search for emission lines of the Ca\,{\sc ii}
triplet near $8600${\AA} and Fe\,{\sc ii} near $5200${\AA}, indicative of the presence of
a gas disk. We used the $4.2$m William Herschel Telescope and the ISIS
spectrograph with the R$=1200$ gratings. No such emission lines were
observed in any object (Figure \ref{mrburleigh01}), although Si\,{\sc iii} absorption was
detected in GD394 (also by \citealp{dupuis00}) and Fe\,{\sc ii} absorption in
HS$0209+083$. 

\section{Discussion - Is there any evidence that hot WDs are accreting
  their metals from disrupted minor planets?}

The non-detection of warm dust or gaseous emission from the 
\citet{barstow03} sample of metal-rich hot WDs is disappointing but not
necessarily a surprise. At a stretch, radiative levitation may still
explain their abundances, although the considerable variations between
objects with similar temperatures and surface gravities suggests
accretion from unseen companions or circumstellar debris (see also
\citealp{vennes06}). The presence of disks around cooler WDs, and the
conclusion that they are accreting the remains of disrupted
terrestrial bodies, suggests the same phenomenon must be occurring at
higher temperatures . Indeed, one well-studied object, GD394
($T_{\rm eff}\approx38\,000$K), 
displays a silicon abundance $\approx100$ times higher than any
other degenerate at similar temperature and gravity. In addition, the
silicon is inhomogeneously distributed across the WD surface ($P_{\rm
  rot} \approx 1.15$~days, \citealp{dupuis00}). There is a strong  temptation to
conclude that GD394 is accreting from surrounding silicate-rich
material, just as with the cooler DAZs, despite the lack of evidence
for the presence of a dust or gas disk in the observations discussed
above. 

In fact, there is evidence for circumstellar material at these hot WDs
in existing far-UV observations. \citet{bannister03} noted the
presence of velocity shifted lines of e.g.\ C\,\,{\sc iv}, Si\,{\sc iv}, and N\,{\sc v} in addition
to photospheric and interstellar components, and discussed possible
origins for these features including ionization of the local
interstellar environment, the presence of material inside the
gravitational well of the white dwarf, mass loss in a stellar wind and
the existence of material in an ancient planetary nebula around the
star. With hindsight, the link to dust and gas disks at cooler DAZs
becomes more enticing. But the question remains - can we convincingly
and conclusively prove that individual hot WDs are accreting from
surrounding material derived from disrupted minor planets in an old
solar system?





\bibliographystyle{aipproc}   


\IfFileExists{\jobname.bbl}{}
 {\typeout{}
  \typeout{******************************************}
  \typeout{** Please run "bibtex \jobname" to optain}
  \typeout{** the bibliography and then re-run LaTeX}
  \typeout{** twice to fix the references!}
  \typeout{******************************************}
  \typeout{}
 }

\end{document}